\documentclass{article}
\usepackage{spconf,amsmath,graphicx}
\usepackage{amsfonts}
\usepackage{mathrsfs}
\usepackage{bbding}

\usepackage{multirow} 
\usepackage{multicol} 
\usepackage{arydshln}
\usepackage{booktabs}
\usepackage{tabu}
\usepackage{hyperref}

\title{Cross-stained Segmentation from Renal Biopsy Images Using Multi-level Adversarial Learning}
\name{Ke Mei$^{\star}$ \qquad Chuang Zhu$^{\star\#}$\thanks{\# the corresponding author: Chuang Zhu (czhu@bupt.edu.cn) }  \qquad Lei Jiang$^{\dagger}$ \qquad Jun Liu$^{\star}$ \qquad Yuanyuan Qiao$^{\star}$ }

\address{$^{\star}$ Center for Data Science, Beijing University of Posts and Telecommunications, Beijing, China\\ $^{\dagger}$ Electron Microscope Lab, Peking University People’s Hospital, Beijing, China}

\begin{document}
%
\maketitle
\begin{abstract}
   Segmentation from renal pathological images is a key step in automatic analyzing the renal histological characteristics. However, the performance of models varies significantly in different types of stained datasets due to the appearance variations. In this paper, we design a robust and flexible model for cross-stained segmentation. It is a novel multi-level deep adversarial network architecture that consists of three sub-networks: (i) a segmentation network; (ii) a pair of multi-level mirrored discriminators for guiding the segmentation network to extract domain-invariant features; (iii) a shape discriminator that is utilized to further identify the output of the segmentation network and the ground truth. Experimental results on glomeruli segmentation from renal biopsy images indicate that our network is able to improve segmentation performance on target type of stained images and use unlabeled data to achieve similar accuracy to labeled data. In addition, this method can be easily applied to other tasks.
\end{abstract}

\section{Introduction}
\label{sec:intro}

\subsection{Background}
\begin{figure}
  \centering 
  \includegraphics[width=3.0in]{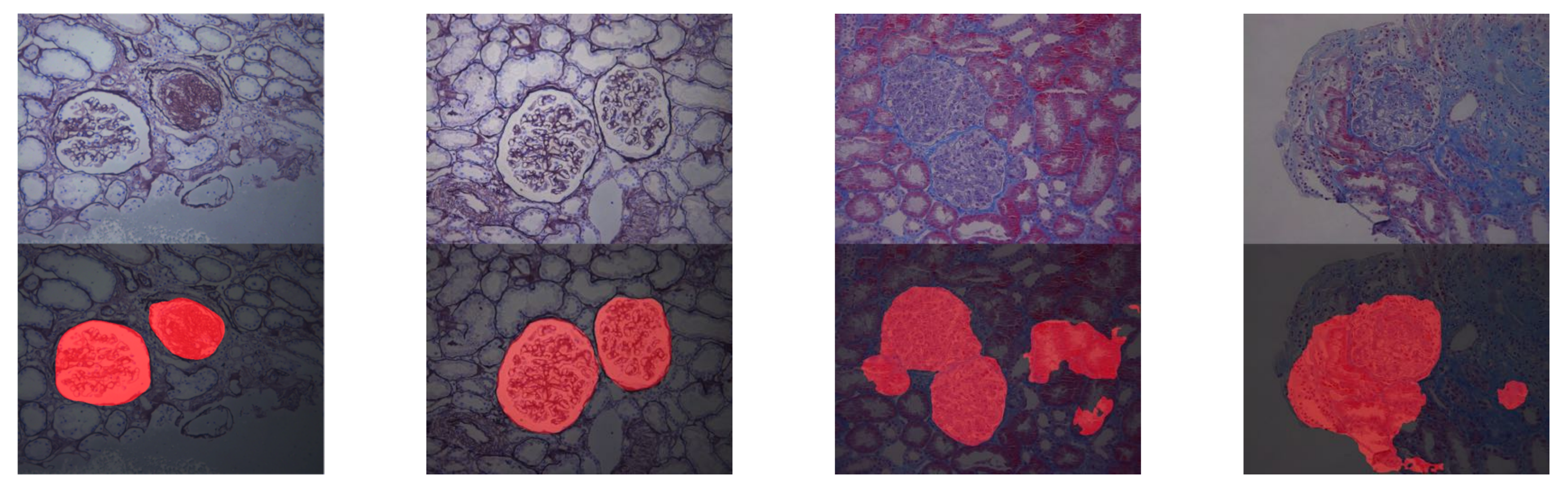}
  \setlength{\belowcaptionskip}{-0.4cm}
  \caption{Illutration of the variations degrading the performance. 
  Row 1: original images, Row 2: prediction by model(trained on PASM), 
  Column 1\&2: images stained by PASM (good), Column 3\&4: images stained by Masson (bad).}\label{fig:1}
  \end{figure}

The histologic examination of renal biopsy slides is of great value to treatment strategies of IgA nephropathy (IgAN) \cite{Rodrigues2017IgA}. Glomeruli segmentation from renal pathological images is a key step in automatic analyzing the renal histological characteristics of IgAN. In clinical practice, histologic examination of renal biopsy tissue requires chemical staining to create contrast. However, for different types of stained images, the complexity of glomeruli segmentation is significantly different. Some staining formulas create additional noise which is not conducive to glomeruli segmentation, such as Masson. In contrast, others enhance the appearance of glomeruli which is conducive to segmentation, such as PASM. Furthermore, due to the scarcity and complexity of pathological images, it is a challenging task to obtain large-scale and finely labeled data. Consequently, we apply knowledge of PASM-stained images to glomeruli segmentation in Masson-stained images in order to improve its performance. However, the appearance variations between two types of stained images can degrade the performance of cross-stain glomeruli segmentation in Figure \hyperref[fig:1]1. In our work, we focus on how to deal with these appearance variations.

\begin{figure*}
  \setlength{\belowcaptionskip}{-0cm}
  \centering
  \includegraphics[width=5.8in]{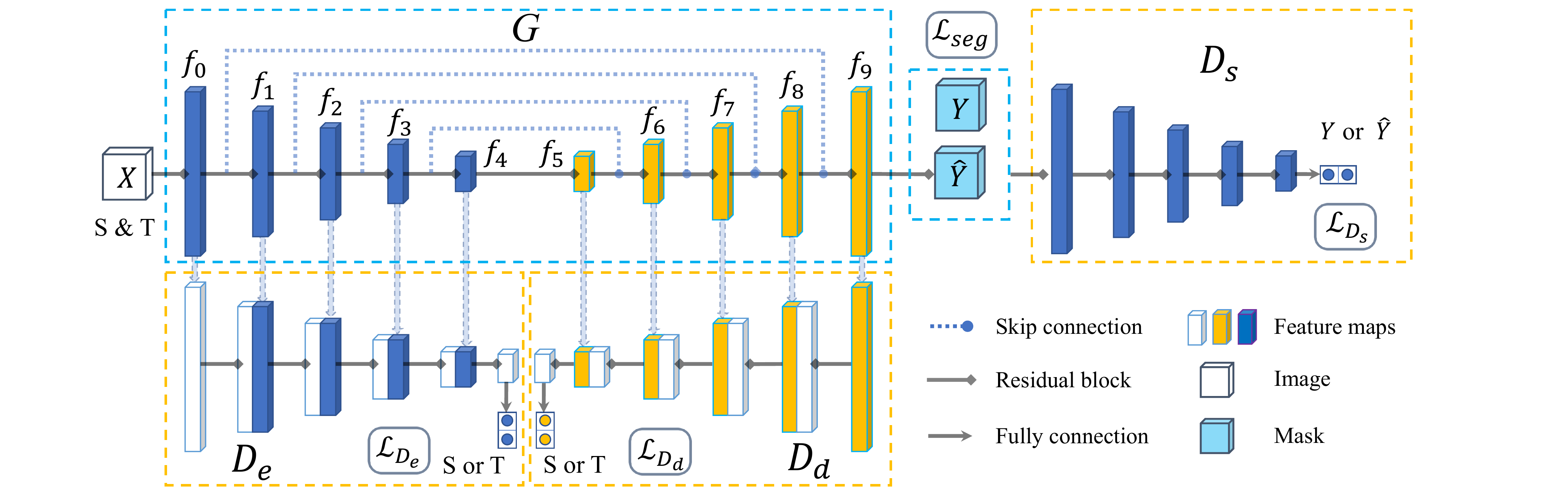}
  \caption{Schematic view of our network architecture. Segmentation network processes the input pathological image to generate a segmentation mask. $D_{e}$, $D_{d}$ determine which domain (S or T) the feature maps in the segmentation network come from. $D_{s}$ distinguishes between the segmentation mask and the ground truth.} \label{fig:2}
\end{figure*}

\subsection{Related Work}
Most of the existing researches address appearance variations by domain adaptation (DA) that assumes the same task with different data distribution between two domains \cite{pan2010survey}. In medical image processing, these appearance variations are addressed by pre-processing. Macenko et al. normalized the stain while retaining the structure \cite{Macenko2009A}. BenTaieb et al. proposed a discriminative image analysis model for stain standardization \cite{bentaieb2018adversarial}. Another approach is to use domain-adversarial networks to impose constraints on the backbone network, allowing the backbone network to learn domain-invariant features. Lafarge et al. proposed a method based on domain-adversarial networks to remove the domain information from the model \cite{lafarge2017domain}. Yang et al. proposed a novel online adversarial appearance conversion solution to explore a composite appearance and structure constraints \cite{yang2018Generalizing}. Dou et al. proposed a unsupervised domain adaptation framework with a domain adaptation module(DAM) and a domain critic module(DCM) \cite{dou2018unsupervised}. Most of the above methods only address DA of a single layer's feature maps (FMs), such as the last layer of the backbone network, but they ignored the information of other layers' FMs. Kamnitsas et al. concatenated the multi-layer FMs after cropping, and then the domain of this concatenated FMs was classified by a domain-adversarial discriminator \cite{kamnitsas2017unsupervised}. However, such concatenated FMs has a huge number of channels and lost information of the cropped FMs, which was not conducive to the discriminator for classification.

\subsection{Relation to Prior Work}
While our work is related to recent approaches \cite{lafarge2017domain,yang2018Generalizing,dou2018unsupervised} in using domain-adversarial networks, we propose a novel multi-level deep adversarial network architecture that includes multiple discriminators for domain adaptation which was not applied in these earlier approaches. In this work, we innovatively propose a pair of mirrored domain-adversarial discriminators for multi-level adversarial learning. We skillfully input the FMs of different layers in the segmentation network into the pair of discriminators, which guides the segmentation network to obtain more domain-invariant features than the earlier approaches \cite{kamnitsas2017unsupervised}. In addition, a shape discriminator further constrains the output of the segmentation network with prior knowledge about shape. Experimental results show that our network can greatly apply knowledge of PASM-stained images to Masson-stained images and improve the performance of segmentation, which can be easily extended to other tasks.

\section{Methodology}
\label{sec:method}
There are two different domains: source domain(S) and target domain(T), which represent two types of stained images. To overcome the domain shift, we use domain-adversarial discriminators that classify which domain the feature of the segmentation network comes from and a shape-adversarial discriminator to constrains the output of the segmentation network. The segmentation network extracts domain-invariant features to trick the discriminator. Our network architecture is shown in Figure \hyperref[fig:2]2. The details are discussed in Section \hyperref[section:2.1]{2.1}.

\subsection{Multi-level Deep Adversarial Network Architecture}\label{section:2.1}

\subsubsection{Segmentation Network.}The segmentation network (G) is the core of our network, which is similar to the generator of GAN \cite{goodfellow2014generative}. Improving the performance of the segmentation network is our ultimate goal. We adopt Unet \cite{ronneberger2015u}, which is widely used in medical image segmentation. It consists of an encoder, a decoder and skip connections. We adopt ResNet-34 \cite{he2016deep} without the last fully connected layer as the encoder. The segmentation network processes the input original images $X$, and obtains a series of FMs $\{f^{(l)}(x)\}$, where $l$ means $l$-th layer. It generates segmentation masks $\hat{Y}$ by these FMs. Finally, we calculate the binary cross entropy of $\hat{Y}$ and labels $Y$ as the initial loss of segmentation network $L_{seg}$.

\subsubsection{Domain-adversarial Discriminators.}
The domain-adversarial discriminator performs binary classification (from S or T) on the FMs in the segmentation network and adversarial train with the segmentation network. In this way, the constrained segmentation network can learn the domain invariant features between S and T. It is intuitive to select the FMs of the last layer to adapt because the FMs of the last layer is more discriminative for the main task. However, in \cite{kamnitsas2017unsupervised}, they found that it is not ideal to only select FMs of the last layer to adapt because the FMs of early layers are more susceptible to appearance variations between domains. 

In \cite{kamnitsas2017unsupervised}, they crop large-sized FMs to match the size of the last layer and concatenate, in order to ensure that all FMs to be concatenated is consistent. However, it will lose a lot of information on the cropped FMs. Moreover, the number of concatenated FMs is too huge, which is difficult for the discriminator to determine the weight of different FMs. The discriminator's attention may be drawn to ‘deep’ features and ‘low’ features may be ignored, which is not conducive to the discriminator for classification. 

Consequently, we skillfully use the network structure (ResNet-34) of the encoder in the segmentation network to mirror a similar network as the encoder discriminator ($D_{e}$). In the encoder part of the segmentation network, the first layer's FMs is input into the discriminator, and the FMs of other layers are sequentially concatenated to the discriminator’s FMs which have the same size as them. In the decoder part of the segmentation network, we also mirror a network as the decoder discriminator ($D_{d}$). Through such a pair of mirrored discriminators ($D_{e}$, $D_{d}$), we can ingeniously solve the problem of the inconsistent size of different layers' FMs instead of cropping the FMs roughly, so that the discriminator is able to use different layers' FMs completely without loss.

We adopt binary cross entropy as loss to update parameters of $D_{e}$ or $D_{d}$. The losses of $D_{e}$ and $D_{d}$ is shown as Eq. \hyperref[equ:1]1 and Eq. \hyperref[equ:2]2, where $p_{S}(x)$ is the distribution of S data, $p_{T}(x)$ is the distribution of T data, $F_{e}(x)=\{f^{(l)}(x)\ |\  l\in encoder\}$ is the FMs of encoder, and $F_{d}(x)=\{f^{(l)}(x)\ |\  l\in decoder\}$ is the FMs of decoder.



\begin{equation}
  \resizebox{1\linewidth}{!}{$\mathcal{L}_{D_{e}}=\mathbb{E}_{x\sim p_{S}(x)}log(D_{e}(F_{e}(x)))+\mathbb{E}_{x\sim p_{T}(x)}log(1-D_{e}(F_{e}(x)))$}
\label{equ:1}
\end{equation}

\begin{equation}
  \resizebox{1\linewidth}{!}{$\mathcal{L}_{D_{d}}=\mathbb{E}_{x\sim p_{S}(x)}log(D_{d}(F_{d}(x)))+\mathbb{E}_{x\sim p_{T}(x)}log(1-D_{d}(F_{d}(x)))$}
\label{equ:2}
\end{equation}

\subsubsection{Shape-adversarial Discriminator.}
In this work, the shape of the segmentation target (glomeruli) is round, and we hope that the shape of predicted mask can be closed to the ground truth. We use this prior knowledge to guide the prediction of the segmentation network, by introducing an additional shape loss, which also contributes to unsupervised domain adaptation. We adopt ResNet-18 as a shape discriminator to achieve this. It distinguishes between the output from the segmentation network and the ground truth. By adversarial learning, it can make the output of the segmentation network as close as possible to the ground truth, thus making their shapes similar. We also adopt binary cross entropy as the loss to update the parameters of $D_{s}$, which is shown as Eq. \hyperref[equ:3]3.

\begin{equation}
  \resizebox{.9\linewidth}{!}{$\mathcal{L}_{D_{s}}=\mathbb{E}_{y\sim p(y)}log(D_{s}(y))+\mathbb{E}_{x\sim p(x)}log(1-D_{s}(G(x)))$}
    \label{equ:3}
\end{equation}

\subsubsection{Combination.}
Combining the ideas presented above, we get the full loss used to update the segmentation network parameters in adversarial training, which is shown as Eq. \hyperref[equ:4]4, where $\alpha_{e}$, $\alpha_{d}$, $\alpha_{s}$ are manually set parameters to balance the weights of four different loss. 

\begin{equation}
  \resizebox{.8\linewidth}{!}{$\mathcal{L}_{full}=\mathcal{L}_{seg}-\alpha_{e} \mathcal{L}_{D_{e}}-\alpha_{d}\mathcal{L}_{D_{d}}-\alpha_{s}\mathcal{L}_{D_{s}}$}
    \label{equ:4}
\end{equation}

\subsection{Training Strategy} \label{section:2.2}
With the images and labels in both S and T, we can train the segmentation network ($G$) and the discriminators ($D_{e}$, $D_{d}$, $D_{s}$) in a supervised way. In the training phase, we try to make $G$ segment more accurately by adapting $\{f^{ (l)}(\cdot )\}$ invariant to variations between S and T. In the initial stage, we train $G$ with $\{\left ( X_{S}, Y_{S} \right ), \left ( X_{T}, Y_{T} \right )\}$  by minimizing $\mathcal{L}_{seg}$, where $\{X_{S}, X_{T}\}$ is the collection of images randomly sampled from S or T, and $\{Y_{S}, Y_{T}\}$ is the collection of their labels. In addition, with the labeled images in S and unlabeled images in T, we can also train $G$ with $\{\left ( X_{S}, Y_{S} \right )\}$ for unsupervised domain adaptation.

After training $G$ for $s_{0}$ epochs, we start to train $D_{e}$, $D_{d}$, $D_{s}$ independently for $d_{0}$ epochs with the trained $G$ by minimizing $\mathcal{L}_{D_{e}}$, $\mathcal{L}_{D_{d}}$ and $\mathcal{L}_{D_{s}}$. 

Then, we obtain a initial $G$ and initial $D_{e}$, $D_{d}$, $D_{s}$, and we start adversarial training them alternately until convergence. In particular, we use $\mathcal{L}_{full}$ as the loss of segmentation network instead of $\mathcal{L}_{seg}$ when training alternately. The experimental results are discussed in Section \hyperref[section:3.2.1]{3.2}.


\begin{table*}[htb]
  \caption{Evaluation of our proposed framework for supervised DA. Mean and standard deviation of Dice coefficient (DC) and accuracy(Acc).}\label{tab1}
  \centering
  \resizebox{1\linewidth}{!}{
  \begin{tabular*}{\hsize}{@{}@{\extracolsep{\fill}}clcccc@{}}
  \toprule
  \multirow{2}{*}{Training Set}                                             & \multicolumn{1}{c}{\multirow{2}{*}{Method}} & \multicolumn{2}{c}{PASM}     & \multicolumn{2}{c}{Masson} \\ \cmidrule(l){3-6} 
                                                                               & \multicolumn{1}{c} {}                        & DC             & Acc          & DC           & Acc          \\ \midrule
  PASM                                                                         & Origin                                       & $93.22\pm0.25$ & $96.36\pm0.14$ & $78.64\pm0.47$ & $94.39\pm0.26$ \\ \midrule
  Masson                                                                       & Origin                                       & $82.31\pm0.31$ & $91.29\pm0.21$ & $86.60\pm0.41$ & $96.22\pm0.13$ \\ \midrule
  \multirow{2}{*}{P \& M} & From scratch                                 & $91.30\pm0.08$   & $95.82\pm0.06$ & $87.08\pm0.26$ & $96.38\pm0.12$ \\
                                                                               & SDA-s                            & $93.70\pm0.16$   & $96.79\pm0.07$ & $90.10\pm0.08$ & $97.15\pm0.05$ \\ 
                                                                               & \textbf{SDA-sed}                             & $\mathbf{94.07\pm0.15}$   & $\mathbf{96.99\pm0.08}$ & $\mathbf{90.49\pm0.34}$ & $\mathbf{97.34\pm0.13}$ \\ \bottomrule
  \end{tabular*}
  }
\end{table*}

\begin{table}[!htb]
  \caption{Evaluation of our proposed framework for unsupervised DA. Mean and standard deviation of Dice coefficient (DC) and accuracy(Acc).}\label{tab2}
  \centering
  \resizebox{.95\linewidth}{!}{
  \begin{tabular*}{\hsize}{@{}@{\extracolsep{\fill}}clcc@{}}
  \toprule
  \multirow{2}{*}{Training Set}                                                          & \multicolumn{1}{c}{\multirow{2}{*}{Method}} & \multicolumn{2}{c}{Test on Masson} \\ \cmidrule(l){3-4} 
                                                                                         & \multicolumn{1}{c}{}                        & DC               & Acc             \\ \midrule
  PASM                                                                                   & Origin                                      & $78.64\pm0.47$     & $94.39\pm0.26$    \\ \midrule
  Masson                                                                                 & Origin                                      & $86.60\pm0.41$     & $96.22\pm0.13$    \\ \midrule    
  \multirow{8}{*}{\begin{tabular}[c]{@{}c@{}}PASM \& \\Masson\\ (Unsuper-\\vised)\end{tabular}} & SF-4                                        & $83.82\pm0.44$     &   $94.52\pm0.16$  \\ 
                                                                                         & SF-9                                   &  $84.95\pm0.34$     &  $94.85\pm0.09$   \\  
                                                                                         & AFC                                   & $85.34\pm0.40$     & $94.91\pm0.14$    \\  \cmidrule(l){2-4}
                                                                                         & UDA-s                                   & $84.27\pm0.34$     & $94.62\pm0.12$    \\  
                                                                                         & UDA-e                                    & $83.97\pm0.37$     & $94.49\pm0.10$    \\ 
                                                                                         & UDA-d                                     & $85.76\pm0.22$     & $93.22\pm0.06$    \\ 
                                                                                         & UDA-ed                               & $87.36\pm0.21$     & $95.54\pm0.10$    \\  
                                                                                         & \textbf{ours}                         & $\mathbf{87.73\pm0.15}$     & $\mathbf{95.59\pm0.06}$    \\ \midrule
  P \& M                                                         
                                                                                         & SDA-sed                         & $90.49\pm0.34$     & $97.34\pm0.13$    \\ \bottomrule
  \end{tabular*}
  }
\end{table}
\section{Experimental Result}
\label{sec:experiment}

\subsection{Materials and Implementation Details}
We used two datasets of renal biopsy pathology images from clinical routines, which are stained respectively with PASM, Masson. They include glomeruli images of normal and multiple lesions, captured at 100x, 200x, and 400x optical Leica Microsystems. Two experienced renal pathologists accurately label the boundaries of the glomerulus in these images. There are variations between the two datasets due to acquisition time, lighting conditions, and chemical staining formulations. 416 images stained with PASM are used as source domain(S), and 403 images stained with Masson are used as target domain(T). Before the experiment, we randomly select 80\% of images from each dataset used for training and the rest for testing.

The proposed method is implemented with the Pytorch 1.0 Framework with a NVIDIA GeForce GTX 1080 Ti. During training, we used the Adam optimizer (initial learning rate is 0.001, momentum parameters $\beta _{1}=0.9$, $\beta _{2}=0.999$) to update the parameters of the networks and set batch size = 4. As mentioned in Section \hyperref[section:2.2]{2.2}, we initially train $G$ for $s_{0}=50$ epochs separately, and then train the discriminator for $d_{0}=10 $ epochs. After that, we alternately train $G$ and discriminators for 100 epochs with $\alpha_{e}=0.01$, $\alpha_{d}=0.05$, $\alpha_{s}=0.1$, which are consistent with supervised DA and unsupervised DA. 

\subsection{Evaluation}\label{section:3.2}
We perform two sets of experiments to verify the superiority of our method compared to baseline in glomeruli segmentation. (i) With T labeled, we use the proposed method for supervised DA and test the performance of the segmentation on both S and T. (ii) With T unlabeled, we use the proposed method for unsupervised DA and test the performance of the segmentation on T.


\paragraph*{Supervised domain adaptation.} \label{section:3.2.1} With T labeled, we evaluate the performance of our method in Table \hyperref[tab1]1. When training with S or T separately, performance degrades due to appearance variations. In the case of scarce data, we use both S and T for domain adaptation (SDA). Adding $D_{s}$ (SDA-s) to the original network can significantly improve the performance. Adding $D_{e}$ and $D_{d}$ (SDA-sed) on the basis of the above can further improve the performance on both S and T. This shows that our method can increase the performance of segmentation on T without losing the accuracy of segmentation on S.
    \paragraph*{Unsupervised domain adaptation.} \label{section:3.2.2}With T unlabeled, we evaluate the performance of our method by transferring S to T in Table \hyperref[tab2]2. We first determine the upper and lower bounds of performance. We use the model trained on S to test directly on T as the lower bound. In contrast, with T labeled, we adopt SDA-sed in Table \hyperref[tab1]1 as the upper bound. In the unsupervised domain adaptation, SF-4 adapts only to the FMs of the single $f_{4}$ layer, SF-9 adapts only to the FMs of the single $f_{9}$ layer, and AFC adapts to the concatenated FMs of all layers. The above three are the baselines of our experiments. Our proposed method (UDA-sed) achieves the best performance, which is very close to the upper bound. We also perform ablation experiments on the methods we proposed. Adding only $D_{e}$(UDA-e) can achieve better performance than SF-4, adding only $D_{d}$ (UDA-d) can achieve better performance than SF-9, and adding $D_{e}$ and $D_{d}$ (UDA-ed) can achieve better results than AFC, which verifies that our ideas are correct.

\section{Conclusion}
We propose a novel multi-level deep adversarial network architecture that includes a segmentation network and multiple adversarial networks for segmenting glomeruli in multi-stained images. We input the feature maps of multiple layers in the segmentation network into the pair of discriminators for adversarial learning, which solves the conflict of inconsistent FMs size without cropping. Experimental results show that our proposed method can greatly improve the performance of glomeruli segmentation in multiple stained images. Moreover, with unlabeled target-stained images, our proposed method can obtain similar performance on labeled target-stained images.

\section{Acknowledgement} This work is supported in part by the Beijing Natural Science Foundation (4182044).

\bibliographystyle{IEEEbib}
\bibliography{strings,refs.bib}

\end{document}